\documentclass[12pt,a4paper]{article}
\pdfoutput=1
\usepackage{jheppub}
\usepackage{tikz}
\usepackage{subcaption}
\usepackage{natbib}
\usepackage{physics}
\usepackage{braket}

\begin{document}

\begin{flushright}
	\hfill{OU-HET-1230}
\end{flushright}
\title{\mathversion{bold} Logarithmic singularities of Renyi entropy as a sign of chaos?}

 \author[a,b,c]{Norihiro Iizuka}
 \author[d]{and Mitsuhiro Nishida}
\affiliation[a]{\it Department of Physics, Osaka University, Toyonaka, Osaka 560-0043, Japan}
\affiliation[b]{\it Department of Physics, National Tsing Hua University, Hsinchu 30013, Taiwan}
\affiliation[c]{\it Yukawa Institute for Theoretical Physics, Kyoto University, Kyoto 606-8502, Japan}
\affiliation[d]{\it Department of Physics, Pohang University of Science and Technology, Pohang 37673, Korea}

\emailAdd{iizuka@phys.nthu.edu.tw, nishida124@postech.ac.kr}

\abstract{We propose that the logarithmic singularities of the Renyi entropy of local-operator-excited states for replica index $n$ can be a sign of quantum chaos. 
As concrete examples, we analyze the logarithmic singularities of the Renyi entropy in various two-dimensional conformal field theories. We show that there are always logarithmic singularities of the Renyi entropy in holographic CFTs, but no such singularities in free and rational CFTs. 
These singularities of the Renyi entropy are also related to the logarithmic time growth of the Renyi entropy at late times. }

\maketitle

\section{Introduction}

It has been proposed that Krylov complexity for operator growth in chaotic quantum mechanical systems grows exponentially in time \cite{Parker:2018yvk}. This exponential growth rate of the Krylov complexity is related to pole-structures of two-point functions on the imaginary time axis. Krylov complexity of operators in quantum mechanical systems has been actively investigated, see for instance, \cite{Barbon:2019wsy, Avdoshkin:2019trj, Dymarsky:2019elm, Rabinovici:2020ryf, Cao:2020zls, Jian:2020qpp, Kim:2021okd, Rabinovici:2021qqt, Caputa:2021sib, Patramanis:2021lkx, Trigueros:2021rwj, Hornedal:2022pkc, Fan:2022xaa, Heveling:2022hth, Bhattacharjee:2022vlt, Muck:2022xfc, Bhattacharya:2022gbz, Rabinovici:2022beu, Liu:2022god, He:2022ryk, Bhattacharjee:2022ave, Bhattacharjee:2022lzy, Alishahiha:2022anw, Bhattacharjee:2023dik, Bhattacharya:2023zqt, Hashimoto:2023swv, Patramanis:2023cwz, Iizuka:2023pov, Bhattacharyya:2023dhp, Camargo:2023eev, Fan:2023ohh, Iizuka:2023fba, Mohan:2023btr, Bhattacharjee:2023uwx, Tang:2023ocr, Bhattacharya:2024uxx, Sasaki:2024puk, Aguilar-Gutierrez:2024nau, Menzler:2024atb, Carolan:2024wov}.

This proposal is problematic in quantum field theories (QFTs) on flat space. Due to continuous degrees of freedom in spatial coordinates, ultraviolet (UV) divergence appears in QFTs. Specifically, two-point functions of operators in QFTs diverge when the two operators are at the same point, whereas there is no such UV divergence in quantum mechanics. For two-point functions in QFTs, inner products between operators shifted in the imaginary time direction are often used. There are theory-independent universal poles in two-point functions on the imaginary time axis due to the UV divergence of QFTs, and hence Krylov complexity grows exponentially in any QFTs. 
Therefore, the exponential growth of Krylov complexity associated with two-point functions is not suitable for a measure of quantum chaos in QFTs \cite{Dymarsky:2021bjq}. For other studies of Krylov complexity in QFTs, see \cite{Magan:2020iac, Kar:2021nbm, Caputa:2021ori, Banerjee:2022ime, Avdoshkin:2022xuw, Camargo:2022rnt, Kundu:2023hbk, Vasli:2023syq, Anegawa:2024wov, Li:2024kfm, Malvimat:2024vhr}.

The reason for the above fault is that we focus on pole-structures of two-point functions that do not depend on the details of QFTs. In fact, the pole-structures of two-point functions due to the UV divergence in QFTs are universally determined up to the conformal dimension of operators if QFTs flow from conformal field theories (CFTs) at UV. To examine quantum chaos in QFTs, we should study quantities that depend on the details of QFTs, and then what quantities should we examine?

In this paper, we propose that the logarithmic singularities of the Renyi entropy for replica index $n$ can be a sign of quantum chaos. Our proposal is motivated by a connection between the exponential growth of Krylov complexity and pole structures of the two-point function as we mentioned. We study the logarithmic singularities of the Renyi entropy of local-operator-excited states in two-dimensional CFTs. We confirm that the logarithmic singularities exist in holographic CFTs but not free or rational CFTs. These logarithmic singularities in holographic CFTs are related to the logarithmic time growth of the Renyi entropy at late times due to the pole-structure of two-point functions.

This paper is organized as follows. In Section \ref{sec:2}, we review the connection between Krylov complexity and singularities of two-point functions and present our proposal for the logarithmic singularities of Renyi entropy as a sign of chaos. In Section \ref{sec:3}, we study explicit examples of the Renyi entropy of local-operator-excited states in two-dimensional CFTs. In Section \ref{sec:4}, the origin of logarithmic singularities in two-dimensional holographic CFTs is explicitly seen. We conclude and discuss future directions in Section \ref{sec:5}.

\section{Our proposal}\label{sec:2}

What we are interested in this paper are quantities that depend on the details of QFTs and are associated with quantum chaos. 
To find a candidate for such quantities, let us consider a thermofield double (TFD) state, which can describe an eternal black hole \cite{Maldacena:2001kr}, with inverse temperature $\beta=1$ unit
\begin{align}
|\text{TFD}\rangle:=\frac{1}{\sqrt{Z}}\sum_j e^{-E_j/2}|E_j\rangle_A|E_j\rangle_B,
\end{align}
where $|E_j\rangle_{A,B}$ are energy eigenstates of subsystems $A$ and $B$ with eigenenergy $E_j$, and $Z:=\sum_j e^{-E_j}$ is a thermal partition function. The reduced density matrix $\rho_A^{\text{TFD}}$ of $|\text{TFD}\rangle$ is a thermal density matrix as
\begin{align}
\rho_A^{\text{TFD}}:=&\,\Tr_B|\text{TFD}\rangle\langle \text{TFD}|\notag\\
=&\,\frac{1}{Z}\sum_j e^{-E_j}|E_j\rangle_A\langle E_j|_A=:e^{-H_A^{\text{TFD}}},\label{rdmTFD}
\end{align}
where we introduce the modular Hamiltonian $H_A^{\text{TFD}}$ of $|\text{TFD}\rangle$. We consider a locally excited state $\mathcal{N}\mathcal{O}_A|\text{TFD}\rangle$ by an operator $\mathcal{O}_A$ on the subsystem $A$, where $\mathcal{N}$ is a normalization factor such that\footnote{In QFTs, we need to introduce a small UV cutoff as done in eq.~(\ref{LOES}).}
\begin{align}
\mathcal{N}^2\langle \text{TFD}|\mathcal{O}_A^\dagger\mathcal{O}_A|\text{TFD}\rangle=1.
\end{align}
Its time evolved state by the Hamiltonian $H_A^{\text{TFD}}\otimes 1_B-1_A\otimes H_B^{\text{TFD}}$ is
\begin{align}
e^{-i(H_A^{\text{TFD}}\otimes 1_B-1_A\otimes H_B^{\text{TFD}})t}\mathcal{N}\mathcal{O}_A|\text{TFD}\rangle.
\end{align}
An inner product between these two states is given by
\begin{align}\label{Ttpf}
&\,\mathcal{N}^2\langle \text{TFD}|\mathcal{O}_A^\dagger e^{-i(H_A^{\text{TFD}}\otimes 1_B-1_A\otimes H_B^{\text{TFD}})t}\mathcal{O}_A|\text{TFD}\rangle\\
=&\,\mathcal{N}^2\langle \text{TFD}|\mathcal{O}_A^\dagger(t)\mathcal{O}_A|\text{TFD}\rangle=\mathcal{N}^2\Tr_A\left[\rho_A^{\text{TFD}}\mathcal{O}_A^\dagger (t)\mathcal{O}_A\right],\notag
\end{align}
which is a thermal two-point function of $\mathcal{O}_A$. Here, we use that the TFD state $|\text{TFD}\rangle$ is invariant under $e^{-i(H_A^{\text{TFD}}\otimes 1_B-1_A\otimes H_B^{\text{TFD}})t}$. If the system is a quantum mechanical system, the pole-structure of this two-point function can be a measure of chaos regarding Krylov complexity of operators \cite{Avdoshkin:2019trj}. However, if the system is a QFT, this two-point function cannot be a measure of chaos as mentioned above.

An explicit example that cannot be a measure of chaos is the entanglement entropy in two-dimensional CFTs on a single interval at finite temperature. In this case, the entanglement entropy can be represented by a two-point function of twist operators, which is universal and cannot be a measure of chaos in CFTs. One way to improve this situation is to consider the entanglement entropy on two disjoint intervals \cite{Hartman:2013mia}, which can be represented by a four-point function of twist operators that depends on the details of CFTs. Another case is four-point out-of-time-order correlators (OTOCs), which can be distinguished between the Ising CFT and holographic CFT cases \cite{Roberts:2014ifa}. The Lyapunov exponent in OTOCs has been proposed as a well-known measure of quantum chaos, and the relationship between the Lyapunov exponent and the exponential growth of Krylov complexity has also been conjectured.

As a further case in this paper, we study the modular time evolution of local-operator-excited states.
Let us consider to modify eq.~(\ref{Ttpf}) in such a way that it depends on the details of QFTs. First consider the Hamiltonian $H_{\mathcal{O}_A}^{\text{TFD}}\otimes 1_B-1_A\otimes H_{\mathcal{O}_{B}}^{\text{TFD}}$ instead of $H_A^{\text{TFD}}\otimes 1_B-1_A\otimes H_B^{\text{TFD}}$, where $H_{\mathcal{O}_A}^{\text{TFD}}$ and $H_{\mathcal{O}_B}^{\text{TFD}}$ are the modular Hamiltonians of $\mathcal{N}\mathcal{O}_A|\text{TFD}\rangle$ defined by
\begin{align}
\rho_{\mathcal{O}}^{\text{TFD}}:=&\,\mathcal{N}^2\mathcal{O}_A|\text{TFD}\rangle\langle \text{TFD}|\mathcal{O}_A^\dagger,\label{rhoTFDO}\\
\rho_{\mathcal{O}_A}^{\text{TFD}}:=&\,\Tr_B\,\rho_{\mathcal{O}}^{\text{TFD}}=:e^{-H_{\mathcal{O}_A}^{\text{TFD}}},\label{rdmTFDOA}\\
\rho_{\mathcal{O}_B}^{\text{TFD}}:=&\,\Tr_A\,\rho_{\mathcal{O}}^{\text{TFD}}=:e^{-H_{\mathcal{O}_B}^{\text{TFD}}}.
\end{align}
We emphasize that $\rho_A^{\text{TFD}}$ in eq. (\ref{rdmTFD}) and $\rho_{\mathcal{O}_A}^{\text{TFD}}$ in eq. (\ref{rdmTFDOA}) are different reduced density matrixes due to the insertion of $\mathcal{O}_A$.  
Since $H_{\mathcal{O}_A}^{\text{TFD}}$ and $H_{\mathcal{O}_B}^{\text{TFD}}$ depend on $\mathcal{O}_A$, this modification could constitute a quantity containing information of higher-point correlation functions of $\mathcal{O}_A$.
However, $H_{\mathcal{O}_A}^{\text{TFD}}\otimes 1_B-1_A\otimes H_{\mathcal{O}_{B}}^{\text{TFD}}$ does not evolve $\mathcal{N}\mathcal{O}_A|\text{TFD}\rangle$ as
\begin{align}\label{tes1}
e^{-i(H_{\mathcal{O}_A}^{\text{TFD}}\otimes 1_B-1_A\otimes H_{\mathcal{O}_B}^{\text{TFD}})t}\mathcal{N}\mathcal{O}_A|\text{TFD}\rangle=\mathcal{N}\mathcal{O}_A|\text{TFD}\rangle.
\end{align}
To evolve the state in a modular time, we next consider an evolution by $H_{\mathcal{O}_A}^{\text{TFD}}\otimes 1_B$. Specifically, the modular time evolution of $\mathcal{N}\mathcal{O}_A|\text{TFD}\rangle$ by $H_{\mathcal{O}_A}^{\text{TFD}}\otimes 1_B$ is given by
\begin{align}\label{mtes}
e^{-i(H_{\mathcal{O}_A}^{\text{TFD}}\otimes 1_B)s}\mathcal{N}\mathcal{O}_A|\text{TFD}\rangle,
\end{align}
where $s$ is the modular time. 
Note that, from eq.~(\ref{tes1}), eq.~(\ref{mtes}) is equivalent to the state evolved by the modular Hamiltonian $1_A\otimes H_{\mathcal{O}_{B}}^{\text{TFD}}$ 
\begin{align}\label{mtes2}
e^{-i(H_{\mathcal{O}_A}^{\text{TFD}}\otimes 1_B)s}\mathcal{N}\mathcal{O}_A|\text{TFD}\rangle=e^{-i(1_A\otimes H_{\mathcal{O}_{B}}^{\text{TFD}})s}\mathcal{N}\mathcal{O}_A|\text{TFD}\rangle.
\end{align}
An inner product between eq.~(\ref{mtes2}) and $\mathcal{N}\mathcal{O}_A|\text{TFD}\rangle$, denoted by $T_{\mathcal{O}}^{\text{TFD}}(s)$, is 
\begin{align}\label{MRATFD}
T_{\mathcal{O}}^{\text{TFD}}(s)=&\,\mathcal{N}^2\langle \text{TFD}|\mathcal{O}_A^\dagger e^{-i(H_{\mathcal{O}_A}^{\text{TFD}}\otimes 1_B)s}\mathcal{O}_A|\text{TFD}\rangle\notag\\
=&\,\mathcal{N}^2\langle \text{TFD}|\mathcal{O}_A^\dagger e^{-i(1_A\otimes H_{\mathcal{O}_{B}}^{\text{TFD}})s}\mathcal{O}_A|\text{TFD}\rangle.
\end{align}
We use subscript $\mathcal{O}$ for $T_{\mathcal{O}}^{\text{TFD}}(s)$ because we focus on the modular time evolution by $H_{\mathcal{O}_A}^{\text{TFD}}\otimes 1_B$ or $1_A\otimes H_{\mathcal{O}_{B}}^{\text{TFD}}$.
This inner product can be interpreted as a two-point function of state $\mathcal{N}\mathcal{O}_A|\text{TFD}\rangle$ that includes information of higher-point correlation functions of $\mathcal{O}_A$ due to the modular Hamiltonians $H_{\mathcal{O}_A}^{\text{TFD}}$ and $H_{\mathcal{O}_B}^{\text{TFD}}$, which depend on the details of QFTs.

The inner product $T_{\mathcal{O}}^{\text{TFD}}(s)$ is a fundamental ingredient of spread complexity, which is a measure of quantum state complexity \cite{Balasubramanian:2022tpr} 
(see also \cite{Caputa:2022eye, Bhattacharjee:2022qjw, Caputa:2022yju, Balasubramanian:2022dnj, Afrasiar:2022efk, Chattopadhyay:2023fob, Erdmenger:2023wjg, Pal:2023yik, Rabinovici:2023yex,  Nandy:2023brt, Gautam:2023bcm, Balasubramanian:2023kwd, Bhattacharya:2023yec, Huh:2023jxt, Beetar:2023mfn, Muck:2024fpb, Zhou:2024rtg}),
 of $\mathcal{N}\mathcal{O}_A|\text{TFD}\rangle$ under the modular time evolution by $1_A\otimes H_{\mathcal{O}_{B}}^{\text{TFD}}$ \cite{Caputa:2023vyr}. 
As the spectral form factor $|Z(\beta+it)|^2$ is examined instead of analytic continued thermal partition function $Z(\beta+it)$, one should focus on $|T_{\mathcal{O}}^{\text{TFD}}(s)|^2$ instead of $T_{\mathcal{O}}^{\text{TFD}}(s)$ by ignoring an overall complex phase factor.
For Krylov complexity, $|T_{\mathcal{O}}^{\text{TFD}}(s)|^2$ is a fundamental ingredient of Krylov complexity of the density matrix operator $\rho_{\mathcal{O}}^{\text{TFD}}$  in eq. (\ref{rhoTFDO}), under the following modular time evolution
\begin{align}
\rho_{\mathcal{O}}^{\text{TFD}}(s):=&\,e^{-i(1_A\otimes H_{\mathcal{O}_{B}}^{\text{TFD}})s}\, \rho_{\mathcal{O}}^{\text{TFD}} \,e^{+i(1_A\otimes H_{\mathcal{O}_{B}}^{\text{TFD}})s}.
\end{align}
To see this, 
one can check \cite{Caputa:2024vrn} that $|T_{\mathcal{O}}^{\text{TFD}}(s)|^2$ can be interpreted as a two-point function of density matrix $\rho_{\mathcal{O}}^{\text{TFD}}$
\begin{align}\label{Ss2}
|T_{\mathcal{O}}^{\text{TFD}}(s)|^2=&\,\Tr\left[\rho_{\mathcal{O}}^{\text{TFD}}(s)\rho_{\mathcal{O}}^{\text{TFD}}\right].
\end{align}

Krylov complexity of operators is a measure of how an initial operator spreads in the Krylov subspace. Krylov basis $\hat{\mathcal{O}}_n$, which is an orthonormal basis defined as
\begin{align}
\Tr\left[\hat{\mathcal{O}}_m^\dagger\hat{\mathcal{O}}_n\right]=\delta_{mn},
\end{align} 
for Krylov complexity associated to $|T_{\mathcal{O}}^{\text{TFD}}(s)|^2$ obeys 
\begin{align}\label{KB1}
\hat{\mathcal{O}}_{-1}:=&\,0, \;\;\; \hat{\mathcal{O}}_0:=\rho_{\mathcal{O}}^{\text{TFD}},\\
[1_A\otimes H_{\mathcal{O}_{B}}^{\text{TFD}}, \hat{\mathcal{O}}_n]=&\,b_n \hat{\mathcal{O}}_{n-1}+b_{n+1} \hat{\mathcal{O}}_{n+1} \;\;\;(n\ge0),\label{EEKS}
\end{align}
where $b_n$ is called the Lanczos coefficient\footnote{From eqs.~(\ref{KB1}) and (\ref{EEKS}), $\hat{\mathcal{O}}_n$ is Hermitian if $n$ is even, and $\hat{\mathcal{O}}_n$ is anti-Hermitian if $n$ is odd. Hence,
one can show that Lanczos coefficient $a_n$ is zero since
\begin{align}
\Tr\left[\hat{\mathcal{O}}_n^\dagger\left[1_A\otimes H_{\mathcal{O}_{B}}^{\text{TFD}}, \hat{\mathcal{O}}_n\right]\right]=(-1)^n\Tr\left[\hat{\mathcal{O}}_n\left[1_A\otimes H_{\mathcal{O}_{B}}^{\text{TFD}}, \hat{\mathcal{O}}_n\right]\right]=0.
\end{align}
}. There is a numerical algorithm to determine $b_n$ from a given $|T_{\mathcal{O}}^{\text{TFD}}(s)|^2$ \cite{RecursionBook}. Eq.~(\ref{EEKS}) represents how $\hat{\mathcal{O}}_n$ spreads to the next basis $\hat{\mathcal{O}}_{n+1}$ by $1_A\otimes H_{\mathcal{O}_{B}}^{\text{TFD}}$, and the Krylov complexity is a measure of this spread in the Krylov subspace. 

As mentioned at the beginning, the exponential growth of Krylov complexity of operators is related to quantum chaos, and its growth rate is determined by the pole-structure of two-point functions on the imaginary time axis. Let us briefly review the exponential growth of Krylov complexity due to the pole of two-point functions on the imaginary axis \cite{Parker:2018yvk}. Suppose that a two-point function $C(s)$ of a Hermitian operator has the closest poles to the origin on the imaginary axis at $s=\pm\frac{i\pi}{2\alpha}$. Then, the spectral function $\Phi(\omega)$, which is defined by the Fourier transformation of $C(s)$, decays exponentially as $\Phi(\omega)\sim e^{-\frac{\pi|\omega|}{2\alpha}}$ at large $|\omega|$ due to the divergence of the following integral at $s=\pm\frac{i\pi}{2\alpha}$
\begin{align}
C(s)=\frac{1}{2\pi}\int_{-\infty}^{+\infty} d\omega\,e^{i\omega s}\, \Phi(\omega).
\end{align}
This exponential decay of $\Phi(\omega)$ leads the asymptotic linear growth of Lanczos coefficient $b_n\sim \alpha n$ at large $n$ \cite{Lubinsky:1988}. Assuming smoothness of $b_n$ with $n$ and the linear growth $b_n\sim \alpha n$, the asymptotic exponential behavior $e^{2\alpha s}$ of Krylov complexity can be derived \cite{Barbon:2019wsy}.

For the Krylov complexity of $\rho_{\mathcal{O}}^{\text{TFD}}$, the two-point function is $|T_{\mathcal{O}}^{\text{TFD}}(s)|^2$ in eq.~(\ref{Ss2}). Therefore, the pole-structure of $T_{\mathcal{O}}^{\text{TFD}}(s)$ with respect to the modular time $s$ is related to chaos under the modular time evolution by $1_A\otimes H_{\mathcal{O}_{B}}^{\text{TFD}}$. We note that the modular Hamiltonian $H_{\mathcal{O}_B}^{\text{TFD}}$ is different from $H_B^{\text{TFD}}$, where $H_B^{\text{TFD}}$ is the Hamiltonian without the insertion of $\mathcal{O}_A$. Hence chaos under the modular time evolution by $1_A\otimes H_{\mathcal{O}_{B}}^{\text{TFD}}$ could be different from usual chaos under the time evolution.

The pole-structure of $T_{\mathcal{O}}^{\text{TFD}}(s)$ is related to the logarithmic singularities of the Renyi entropy. Moreover, this relation can be generalized to a general pure state $|\Psi\rangle$. To see these, we review the relationship between the Renyi entropy and $T_{\mathcal{O}}^{\text{TFD}}(s)$ \cite{Caputa:2023vyr}. Let us consider a pure state $|\Psi\rangle$, as a generalization of $\mathcal{N}\mathcal{O}_A|\text{TFD}\rangle$, on a total Hilbert space $\mathcal{H}$ and assume that $\mathcal{H}$ can be decomposed into two Hilbert spaces as $\mathcal{H}=\mathcal{H}_A\otimes\mathcal{H}_B$ for two subsystems $A$ and $B$.\footnote{In QFTs, this decomposition may not hold in general. In this paper, we assume this decomposition by introducing UV cutoff.} Schmidt decomposition of $|\Psi\rangle$ is given by 
\begin{align}
|\Psi\rangle=\sum_j \sqrt{\lambda_j}|j\rangle_A|j\rangle_B,
\end{align}
where $|j\rangle_A$ and $|j\rangle_B$ are basis states in $\mathcal{H}_A$ and $\mathcal{H}_B$, respectively.  The reduced density matrix $\rho_{\mathcal{O}_B}$ for the subsystem $B$ and its modular Hamiltonian $H_{\mathcal{O}_B}$ are 
\begin{align}\label{HA}
\rho_{\mathcal{O}_B}:=\Tr_A|\Psi\rangle\langle \Psi|=\sum_j \lambda_j|j\rangle_B\langle j|_B=:e^{-H_{\mathcal{O}_B}},
\end{align}
where we use subscript $\mathcal{O}_B$ to keep in mind that we will consider a locally excited state as $|\Psi\rangle$ and its reduced density matrix for the subsystem $B$.
From this reduced density matrix $\rho_{\mathcal{O}_B}$, the Renyi entropy $S_{\mathcal{O}_B}^{(n)}$ is defined by
\begin{align}\label{RE1}
S_{\mathcal{O}_B}^{(n)}:=\frac{1}{1-n}\log \Tr_B (\rho_{\mathcal{O}_B})^n=\frac{1}{1-n}\log\sum_j\lambda_j^{n},
\end{align}
where $n$ is the replica index.

By using the modular Hamiltonian $1_A\otimes H_{\mathcal{O}_{B}}$, we define the modular time evolution of $|\Psi\rangle$ as
\begin{align}
|\Psi(s)\rangle:=e^{-i(1_A\otimes H_{\mathcal{O}_{B}})s}| \Psi \rangle.
\end{align}
 As an inner product between $|\Psi(s)\rangle$ and $|\Psi\rangle$, the modular return amplitude $T_{\mathcal{O}}(s)$ with $1_A\otimes H_{\mathcal{O}_{B}}$ is defined by
\begin{align}\label{MRA1}
T_{\mathcal{O}}(s):=\langle \Psi|\Psi(s)\rangle=\langle \Psi |e^{-i(1_A\otimes H_{\mathcal{O}_{B}})s}|\Psi\rangle=\sum_j\lambda_j^{1+is},
\end{align}
where we use that eigenvalues of $e^{-H_{\mathcal{O}_B}}$ are $\lambda_j$. This modular return amplitude $T_{\mathcal{O}}(s)$ is a generalization of $T_{\mathcal{O}}^{\text{TFD}}(s)$ in eq. (\ref{MRATFD}).
From eqs.~(\ref{RE1}) and (\ref{MRA1}) with an analytic continuation of the replica index $n\to1+is$, we obtain the desired relationship
\begin{align}\label{RelationRenyiSs}
S_{\mathcal{O}_B}^{(1+is)}=&\frac{i}{s}\log T_{\mathcal{O}}(s).
\end{align}

From this relation, logarithmic singularities of $S_{\mathcal{O}_B}^{(1+is)}$ with respect to $s$ come from poles of $T_{\mathcal{O}}(s)$. Therefore, the Renyi entropy $S_{\mathcal{O}_B}^{(1+is)}$ has logarithmic singularities on the imaginary axis of $s$ if the modular time evolution by $1_A\otimes H_{\mathcal{O}_{B}}$ is chaotic, where $H_{\mathcal{O}_B}$ is given by eq. (\ref{HA}).
Furthermore, we propose that the chaotic nature by the modular Hamiltonian $1_A\otimes H_{\mathcal{O}_{B}}$ is related to the chaotic nature of QFTs. In the rest of the paper, we study several two-dimensional CFT examples and see there are indeed connections between them.
Note that the singularities on the imaginary axis of $s$ are equivalent to the singularities on the real axis of the replica index $n=1+is$.

\section{Two-dimensional CFT examples}\label{sec:3}

We follow the setup in \cite{Nozaki:2014hna, Caputa:2014vaa} for the Renyi entropy of local-operator-excited states in two-dimensional CFTs. We consider a two-dimensional Euclidean space with space coordinate $x$, Euclidean time coordinate $\tau$, and its analytic continuation to Lorentzian time $t$ as follows. 
Let $\mathcal{O}(0,x_1)$ be a local operator at $\tau=0$ and $x=x_1$. 
Let us consider the evolution of this operator into Euclidean and Lorentzian time as follows
\begin{align}
\mathcal{O}(\tau_1,x_1)=&\,e^{H\tau_1}\mathcal{O}(0,x_1)e^{-H\tau_1},\\
\mathcal{O}(\tau_1+it_1,x_1)=&\, e^{iHt_1}\mathcal{O}(\tau_1,x_1)e^{-iHt_1},
\end{align}
where $H$ is the Hamiltonian of CFTs.
Since we consider the analytic continuation to real time, $\mathcal{O}(\tau_1+it_1,x_1)$ depends on three parameters $\tau_1,t_1,x_1$. 

Let us consider a pure state
\begin{align}\label{LOES}
|\Psi\rangle=&\,\mathcal{N}e^{-iHt}e^{-\epsilon H}\mathcal{O}(0,-l)|0\rangle=\mathcal{N}\mathcal{O}(-\epsilon-it,-l)|0\rangle.
\end{align}
Here, $|0\rangle$ is the ground state of CFTs, where $H|0\rangle=0$, and $\mathcal{O}(0,-l)$ is an operator located at $\tau=0$ and $x=-l<0$. We introduce a small UV cutoff $\epsilon$, and $\mathcal{N}$ is a normalization factor such that $\langle \Psi|\Psi\rangle=1$. 
This pure state $|\Psi\rangle$ depends on time $t$, and we call $t\gg l\gg \epsilon$ as late time limit. Note that this Lorentzian time $t$ is different from the modular time $s$. From now, we consider the chaotic nature of the modular time evolution of this $|\Psi\rangle$ and logarithmic singularities of the Renyi entropy in the late time limit $t\gg l\gg \epsilon$.

The density matrix $\rho_{\mathcal{O}}=|\Psi\rangle\langle \Psi|$ of this pure state is given by
\begin{align}
\rho_{\mathcal{O}}=&\,\mathcal{N}^2e^{-iHt}e^{-\epsilon H}\mathcal{O}(0,-l)|0\rangle\langle0|\mathcal{O}^\dagger(0,-l)e^{-\epsilon H} e^{+iHt}\notag\\
=&\,\mathcal{N}^2\mathcal{O}(-\epsilon-it,-l)|0\rangle\langle0|[\mathcal{O}(-\epsilon-it,-l)]^\dagger\notag\\
=&\,\mathcal{N}^2\mathcal{O}(-\epsilon-it,-l)|0\rangle\langle0|\tilde{\mathcal{O}}^\dagger(\epsilon-it,-l),
\end{align}
where our notation of Hermitian conjugate is
\begin{align}
&\,[\mathcal{O}(-\epsilon-it,-l)]^\dagger=[e^{-\epsilon H}\mathcal{O}(-it,-l)e^{+\epsilon H}]^\dagger\notag\\
=&\,e^{+\epsilon H}\mathcal{O}^\dagger(-it,-l)e^{-\epsilon H}=:\tilde{\mathcal{O}}^\dagger(\epsilon-it,-l).
\end{align}
Note that $\tilde{\mathcal{O}}^\dagger(\epsilon-it,-l)$ is defined by the Euclidean time $\epsilon$ evolution of $\mathcal{O}^\dagger(-it,-l)$, thus the sign of $\epsilon$ is flipped. 

From now, we write $\mathcal{O}(-\epsilon-it,-l)$ as $\mathcal{O}(w,\bar{w})$ by defining
\begin{align}
\mathcal{O}(w,\bar{w}):=\mathcal{O}(-\epsilon-it,-l),
\end{align}
where 
\begin{align}
w:=+i(-\epsilon-it)-l, \;\;\; \bar{w}:=-i(-\epsilon-it)-l.
\end{align}
Note that $\bar{w}$ is not complex conjugate of $w$ due to $t$.
With this notation, the density matrix $\rho_{\mathcal{O}}$ can be expressed as
\begin{align}
\rho_{\mathcal{O}}=\mathcal{N}^2\mathcal{O}(w_2,\bar{w}_2)|0\rangle\langle0|\tilde{\mathcal{O}}^\dagger(w_1,\bar{w}_1),
\end{align}
where new coordinates $w_i$ are given by
\begin{align}
\begin{split}
w_1=&\,+i(+\epsilon-it)-l, \;\;\; \bar{w}_1=-i(+\epsilon-it)-l,\\
w_2=&\,+i(-\epsilon-it)-l, \;\;\; \bar{w}_2=-i(-\epsilon-it)-l.\label{w1w2}
\end{split}
\end{align}
We choose subsystem $A$ to be a half space $x<0$ in negative direction and subsystem $B$ to be a half space $x>0$ in positive direction. All of our operators are inserted on the subsystem $A$. Then, the reduced density matrix $\rho_{\mathcal{O}_B}$ for the subsystem $B$ is defined by $\rho_{\mathcal{O}_B}=\Tr_{A}\,\rho_{\mathcal{O}}=e^{-H_{\mathcal{O}_B}}$.

We define the difference of the Renyi entropy $\Delta S_B^{(n)}$ for the subsystem $B$ between the excited state $|\Psi\rangle$ in eq.~(\ref{LOES}) and the ground state $|0\rangle$ as
\begin{align}
\Delta S_B^{(n)}:=&\,\frac{1}{1-n}\log \Tr \rho_{\mathcal{O}_B}^n-\frac{1}{1-n}\log \Tr \rho_B^n,\notag\\
=&\,\frac{1}{1-n}\log \frac{Z_n}{(Z_1)^n}-\frac{1}{1-n}\log \frac{Z^{(0)}_n}{(Z^{(0)}_1)^n},
\end{align}
where $\rho_B$ is the reduced density matrix of the ground state $|0\rangle$. Here, $Z_n$ and $Z_n^{(0)}$ are the replica partition functions on $\Sigma_n$ with and without the insertion of $\mathcal{O}$, respectively, where $\Sigma_n$ is the $n$-sheeted replica geometry constructed from $n$ copies of $\Sigma_1=\mathbb{R}^2$ by gluing them at the subsystem $B$. 
The reason for taking this difference $\Delta S_B^{(n)}$ is to focus on the effect of the modular Hamiltonian $H_{\mathcal{O}_B}$ with local excitation by $\mathcal{O}$ on the Renyi entropy, and we analyze logarithmic singularities of $\Delta S_B^{(n)}$.
By using the replica method, one can express $\Delta S_B^{(n)}$ by correlation functions of $\mathcal{O}$ as
\begin{align}
\Delta S_B^{(n)}\label{SAn}
=&\,\frac{1}{1-n}\log \frac{Z_n}{Z_n^{(0)}}-\frac{1}{1-n}\log \frac{(Z_1)^n}{(Z^{(0)}_1)^n}\notag\\
=&\,\frac{1}{1-n}\log \frac{\langle\tilde{\mathcal{O}}^\dagger(w_1,\bar{w}_1)\mathcal{O}(w_2,\bar{w}_2)\cdots \mathcal{O}(w_{2n},\bar{w}_{2n}\rangle_{\Sigma_n}}{\left(\langle\tilde{\mathcal{O}}^\dagger(w_1,\bar{w}_1)\mathcal{O}(w_2,\bar{w}_2)\rangle_{\Sigma_1}\right)^n}.
\end{align}

Coordinates $w$ and $\bar{w}$ on $\Sigma_n$ can be mapped to  coordinates $z$ and $\bar{z}$ on $\Sigma_1$ as
\begin{align}\label{CM}
z=w^{1/n}=e^{(\phi+i\tau)/n}, \;\;\; \bar{z}=\bar{w}^{1/n}=e^{(\phi-i\tau)/n}.
\end{align}
We can express positions of operators by using coordinates 
\begin{align}\label{wphitau}
w_j=e^{(\phi_j+i\tau_j)}, \;\;\; \bar{w}_j=e^{(\phi_j-i\tau_j)},
\end{align}
with the following periodicity 
\begin{align}
\phi_{2k+1}=&\phi_1,\;\;\; \phi_{2k+2}= \phi_2,\\
\tau_{2k+1}=&\tau_1+2\pi k, \;\;\; \tau_{2k+2}=\tau_2+2\pi k,
\end{align}
where $k+1$ is the label of replica sheet. Note that inserting operators at different locations of $\tau_j$ is similar to shifting the coordinates of operators  in OTOCs by $\beta/2$ in the imaginary time direction.
Here, $\phi_j$ and $\tau_j$ for $j=1,2$ can be determined by eq. (\ref{w1w2}),
where we choose\footnote{In Appendix \ref{AppA}, we explain an explicit method to determine $\phi_j$ and $\tau_j$ for $j=1,2$.}
\begin{align}\label{condition}
0\le\Im [\phi_j+i\tau_j]\le2\pi, \;\;\; -2\pi\le\Im [\phi_j-i\tau_j]\le0.
\end{align}
If $t\neq 0$, since $\bar{z}_j$ is not complex conjugate of $z_j$, $\phi_j$ and $\tau_j$ are generally complex. One can confirm that $\Delta \phi:=\phi_2-\phi_1$ is pure imaginary, and $\Delta \tau:=\tau_2-\tau_1$ is real.

In the late time limit, one can obtain
\begin{align}
\Im(\phi_1+i\tau_1)\sim&\,0, \;\;\;\;\;\Im(\phi_1-i\tau_1)\sim-\pi,\\
\Im(\phi_2+i\tau_2)\sim&\,2\pi, \;\;\; \Im(\phi_2-i\tau_2)\sim-\pi
\end{align}
and therefore
\begin{align}\label{DpDtLTL}
\Delta\phi\sim i\pi, \;\;\; \Delta\tau\sim\pi.
\end{align}

Late-time behavior of $\Delta S_B^{(n)}$ with $t\gg l\gg \epsilon$ in various CFTs has been studied by analyzing the correlation function $\langle\tilde{\mathcal{O}}^\dagger(w_1,\bar{w}_1)\mathcal{O}(w_2,\bar{w}_2)\cdots \mathcal{O}(w_{2n},\bar{w}_{2n}\rangle_{\Sigma_n}$ \cite{He:2014mwa, Nozaki:2014uaa, Caputa:2014vaa, Nozaki:2014hna, Asplund:2014coa, Caputa:2014eta, Guo:2015uwa, Caputa:2015waa, Chen:2015usa, Numasawa:2016kmo, Caputa:2017tju, Guo:2018lqq, Kusuki:2019gjs,  Caputa:2019avh, Kusuki:2019avm}. We summarize some typical results among them. 

\subsection{Example 1: free massless scalar}
Let $\varphi$ be a free massless scalar in two-dimension, and consider 
\begin{align}
\mathcal{O}=:e^{ip\varphi}:+:e^{-ip\varphi}:,
\end{align}
where $p$ is a real constant, and $:\;:$ represents normal ordering. The late-time behavior of $\Delta S_B^{(n)}$ with this operator is \cite{Nozaki:2014hna}
\begin{align}\label{SAFreeCFTs}
\Delta S_B^{(n)}\sim \log 2,
\end{align}
which does not depend on $n$. Therefore, $\Delta S_B^{(n)}$ at late times does not have logarithmic singularities with respect to $n$. Similar computations of free massless scalars in higher-dimension were done in \cite{Nozaki:2014hna,  Nozaki:2014uaa}.

\subsection{Example 2: rational CFTs}
We consider two-dimensional rational CFTs and a primary operator $\mathcal{O}_a$, where $a$ is the label of primary operators. The late-time behavior of $\Delta S_B^{(n)}$ is given by \cite{He:2014mwa}
\begin{align}\label{SARationalCFTs}
\Delta S_B^{(n)}\sim \log d_a,
\end{align}
where $d_a$ is called quantum dimension that satisfies
\begin{align}
d_a=\frac{1}{F_{00}[a]}=\frac{S_{0a}}{S_{00}},
\end{align} 
where $F_{00}[a]$ and $S_{ab}$ are Fusion matrix and modular $S$-matrix in rational CFTs, respectively \cite{Moore:1988uz, Verlinde:1988sn, Moore:1988ss}. Here, the index $0$ represents the identity operator. As in the case of free massless scalar, $\Delta S_B^{(n)}$ at late times is $n$-independent and does not have logarithmic singularities.

\subsection{Example 3: holographic CFTs}
The authors of \cite{Caputa:2014vaa} studied the late-time behavior of the Renyi entropy in two-dimensional holographic CFTs with large central charge $c$, where $\mathcal{O}$ is a scalar primary operator with chiral conformal dimension $\Delta_{\mathcal{O}}/c\ll1$. Under the large $c$ approximation, the late-time behavior of $\Delta S_B^{(n)}$ was derived as
\begin{align}
\Delta S_B^{(n)}\sim&\, \frac{1}{1-n}\log\left[2\left(\frac{\epsilon}
{nt \sin\left(\frac{\pi}{n}\right)}\right)^{2n\Delta_\mathcal{O}}\right]
\notag\\
=&\,\frac{2n\Delta_\mathcal{O}}{n-1}\log\left[\frac{nt\sin\left(\frac{\pi}{n}\right)}{\epsilon}\right]-\frac{1}{n-1}\log 2.\label{SAHolographicCFTs}
\end{align}
Note the order of the limits, we first take (1) large $c$ limit, then (2) late time limit $t\gg l\gg \epsilon$.
This expression of the Renyi entropy includes $\log\left[\sin\left(\frac{\pi}{n}\right)\right]$, which is quite different from the free and rational CFTs above. With the analytic continuation $n\to1+is$, $\sin\left(\frac{\pi}{n}\right)=\sin\left(\frac{\pi}{1+is}\right)$ can be zero on the imaginary axis of $s$. Therefore, in two-dimensional holographic CFTs with large $c$, the Renyi entropy $\Delta S_B^{(1+is)}$ for local-operator-excited states has logarithmic singularities on the imaginary axis of $s$. This result implies a connection between the chaotic nature of modular Hamiltonian and the chaotic nature of two-dimensional holographic CFTs. Later, we show that there are always logarithmic singularities in holographic CFTs even though we do not take the late time limit.

In the bulk picture, the above Renyi entropy can be computed from a geodesic length on the following three-dimensional Euclidean bulk geometry 
\begin{align}\label{AdS3BH}
ds^2=f(r)d\tau^2+\frac{dr^2}{f(r)}+r^2d\phi^2, \;\;\; f(r)=r^2-\frac{1}{n^2},
\end{align}
where $\phi\in\mathbb{R}$ is non-compact, and the periodicity of $\tau$ is $2\pi n$. These $\phi$ and $\tau$ correspond to boundary coordinates in eq.~(\ref{CM}) at $t=0$.
For simplicity, we set the AdS radius as $R=1$. 
The geodesic length between two points on the AdS boundary can be written by a function of $\Delta \phi$ and $\Delta \tau$, which are real variables in Euclidean signature. By using the analytic continuation to complex variables with nonzero $t$, one can calculate the Renyi entropy from the geodesic length, which agrees with eq.~(\ref{SAHolographicCFTs}) \cite{Caputa:2014vaa}.

\section{Origin of $\log\left[\sin\left(\frac{\pi}{n}\right)\right]$ in two-dimensional holographic CFTs} \label{sec:4}

Let us see how $\log\left[\sin\left(\frac{\pi}{n}\right)\right]$ appears in two-dimensional holographic CFTs by reviewing the derivation of eq.~(\ref{SAHolographicCFTs}) \cite{Caputa:2014vaa}. The key property of large $c$ limit in holographic CFTs is the factorization,  namely $2n$-point function $\langle\tilde{\mathcal{O}}^\dagger(w_1,\bar{w}_1)\mathcal{O}(w_2,\bar{w}_2)\cdots \mathcal{O}(w_{2n},\bar{w}_{2n}\rangle_{\Sigma_n}$ factorizes to products of two-point functions. By using the conformal map in eq.~(\ref{CM}),
a two-point function on $\Sigma_n$ is given by
\begin{align}
&\,\langle\tilde{\mathcal{O}}^\dagger(w_1,\bar{w}_1)\mathcal{O}(w_2,\bar{w}_2)\rangle_{\Sigma_n}\notag\\
=&\, n^{-4\Delta_{\mathcal{O}}}|w_1w_2|^{\frac{2\Delta_\mathcal{O}(1-n)}{n}}|w_1^{1/n}-w_2^{1/n}|^{-4\Delta_{\mathcal{O}}},
\end{align}
thus,
\begin{align}
&\,\frac{\langle\tilde{\mathcal{O}}^\dagger(w_1,\bar{w}_1)\mathcal{O}(w_2,\bar{w}_2)\rangle_{\Sigma_n}}{\langle\tilde{\mathcal{O}}^\dagger(w_1,\bar{w}_1)\mathcal{O}(w_2,\bar{w}_2)\rangle_{\Sigma_1}}\notag\\
=&\,\left(\frac{\cosh (\Delta \phi)-\cos (\Delta \tau)}{n^2(\cosh (\Delta \phi/n)-\cos (\Delta \tau/n))}\right)^{2\Delta_{\mathcal{O}}}.\label{tpf}
\end{align}

There are various choices of Wick contractions of the $2n$-point function. In the late time limit, two of them are dominant. First one is the contraction of operators on the same replica sheet, where $\Delta \phi=\phi_2-\phi_1$ and $\Delta\tau=\tau_2-\tau_1$ are given by eq.~(\ref{DpDtLTL}). Second one is the contraction of operators on the $k$-th sheet and on the $k+1$-th sheet, where $\Delta \phi=\phi_3-\phi_2$ and $\Delta\tau=\tau_3-\tau_2$ are given by 
\begin{align}
\Delta\phi\sim -i\pi, \;\;\; \Delta\tau\sim\pi.
\end{align}
Such Wick contractions are dominant contributions because $\cosh (\Delta \phi/n)-\cos (\Delta \tau/n)\sim0$ holds in the denominator in eq.~({\ref{tpf}). Thus, the others are suppressed by $\epsilon/t$. 
 Therefore, in the late time limit, the $2n$-point function can be approximated by the above two types of Wick contractions as products of two-point functions
\begin{align}\label{factorized2n}
&\,\frac{\langle\tilde{\mathcal{O}}^\dagger(w_1,\bar{w}_1)\mathcal{O}(w_2,\bar{w}_2)\cdots \mathcal{O}(w_{2n},\bar{w}_{2n}\rangle_{\Sigma_n}}{\left(\langle\tilde{\mathcal{O}}^\dagger(w_1,\bar{w}_1)\mathcal{O}(w_2,\bar{w}_2)\rangle_{\Sigma_1}\right)^n}\notag\\
\sim&\, 2 \left(\frac{\langle \tilde{\mathcal{O}}^\dagger(w_1,\bar{w}_1)\mathcal{O}(w_2,\bar{w}_2)\rangle_{\Sigma_n}}{\langle \tilde{\mathcal{O}}^\dagger(w_1,\bar{w}_1)\mathcal{O}(w_2,\bar{w}_2)\rangle_{\Sigma_1}}\right)^n.
\end{align}

From eqs.~(\ref{SAn}) and (\ref{factorized2n}), logarithmic singularities of $\Delta S_B^{(n)}$ comes from poles of the two-point function in eq.~(\ref{tpf}) with respect to $n$. Specifically, the poles are determined by
\begin{align}\label{poleEq}
\cosh (\Delta \phi/n)-\cos (\Delta \tau/n)=0.
\end{align}
Since $\Delta\phi$ is pure imaginary and $\Delta\tau$ is real, the solutions of eq.~(\ref{poleEq}) are  
\begin{align}
\frac{\Delta\tau}{n}=\pm i\frac{\Delta\phi}{n}+2\pi k,
\end{align}
where $k$ is an integer. This gives 
\begin{align}\label{polen}
n=\frac{\Delta \tau\pm i\Delta\phi}{2\pi k}.
\end{align}
Thus, for given $\Delta\phi$ and $\Delta\tau$, there are always poles at real values of $n$ given by eq.~(\ref{polen}).

In the late time limit $\frac{t}{\epsilon}\to\infty$, by using eq.~(\ref{eqphitau}) in Appendix \ref{AppA}, one can evaluate\footnote{If $\sin (\pi/n)$ is exactly zero, we must consider the contribution from $O(t^{-2})$ terms. In Appendix \ref{AppB}, we numerically check that eq.~(\ref{poleEq}) holds in the late time limit if $\sin (\pi/n)\sim0$.}
\begin{align}\label{CoshPhin-CosTaun}
\cosh (\Delta \phi/n)-\cos (\Delta \tau/n)
=&\,\frac{2\epsilon}{nt} \sin (\pi/n)+O(t^{-2}),
\end{align}
and this is the origin of $\sin\left(\frac{\pi}{n}\right)$ in eq.~(\ref{SAHolographicCFTs}). 
This is just the result of eq.~(\ref{polen}) with late time limit eq.~(\ref{DpDtLTL}).
In holographic CFTs, where $c\to\infty$, the $2n$-point functions factorize into products of two-point functions. Even though there are many ways of Wick contractions, note that generically there is a dominant Wick contraction. And for that one, just as we showed in eqs.~(\ref{factorized2n}), (\ref{poleEq}), (\ref{polen}), one can find poles at real values of $n$.

\section{Conclusion}\label{sec:5}

We have pointed out a new connection between quantum chaos under the modular time evolution and the logarithmic singularities of the Renyi entropy for replica index $n$. This connection is motivated by the exponential growth of the Krylov complexity of the density matrix operator under the modular time evolution. We have studied the logarithmic singularities in several two-dimensional CFT examples and found a connection between the chaotic nature of the modular Hamiltonian and the chaotic nature of CFTs.

Two-point functions are universal in any two-dimensional CFTs, therefore these by themselves are not enough as measures of chaos in CFTs. However, in holographic CFTs, since $2n$-point functions factorize into products of two-point functions, the logarithmic singularities of the Renyi entropy $\Delta S_B^{(n)}$ for $n$ come from the poles of the two-point function in eq.~(\ref{tpf}). We showed that, in holographic CFTs, there are always poles given by eq.~(\ref{polen}) at real values of $n$. These poles result in the logarithmic singularities of the Renyi entropy in holographic CFTs.

From the viewpoint of time evolution in $t$, $\Delta S_B^{(n)}$ in two-dimensional holographic CFTs in eq.~(\ref{SAHolographicCFTs})  grows logarithmically as $\log t$, but $\Delta S_B^{(n)}$ in free (eq.~(\ref{SAFreeCFTs})) and rational (eq.~(\ref{SARationalCFTs})) CFTs becomes a constant at late times. The logarithmic growth of the Renyi entropy in time $t$ seems to be related to the chaotic property of holographic CFTs \cite{Asplund:2014coa, Caputa:2014vaa, Kusuki:2019gjs}.
In two-dimensional CFTs, our discussion connecting chaos by the modular Hamiltonian and the logarithmic singularities of the Renyi entropy for replica index $n$ is related to the above expectation of the logarithmic growth in $t$, since the logarithmic growth in $t$ and the logarithmic singularities in $n$ from $\log\left[\sin\left(\frac{\pi}{n}\right)\right]$ are connected due to eq.~(\ref{CoshPhin-CosTaun}).

There are many other Renyi generalizations of quantum information measures such as Renyi mutual information, Renyi divergence, and Renyi reflected entropy. It is interesting to study and formulate connections between their singular structures in the replica index $n$ and quantum chaos. It is also interesting to investigate the connection we discussed in higher-dimensional settings.

Finally, we comment on the generalization of our calculations to local-operator-excited states at finite temperature. The time evolution of Renyi entropy of local-operator-excited states in two-dimensional CFTs at finite temperature was studied by \cite{Caputa:2014eta}. Since they only gave the results of Renyi entropy at $n=2$, it is significant to generalize their results to general values of $n$ and evaluate the logarithmic singularities in $n$. One of their key conclusions at finite temperature is that the time evolution of Renyi entropy at $n=2$ in the large $c$ limit stops at $t\sim\beta$, and the Renyi entropy saturates to a value $\Delta S^{(2)}\sim\Delta_\mathcal{O}\log \frac{\beta}{\epsilon}$, where $\Delta_{\mathcal{O}}/c\ll1$. This $\beta$-dependent time scale reminds us of the dissipation time $t_d\sim\beta$ and the scrambling time $t_*\sim\frac{\beta}{2\pi}\log c$ in OTOCs at finite temperature \cite{Roberts:2014ifa, Maldacena:2015waa}. It would be interesting to closely examine the logarithmic singularities in the approximate expressions of Renyi entropy within a specific time range, whether at early times or late times compared to $\beta$.

\acknowledgments
The work of NI was supported in part by JSPS KAKENHI Grant Number 18K03619, MEXT KAKENHI Grant-in-Aid for Transformative Research Areas A “Extreme Universe” No. 21H05184. M.N. was supported by the Basic Science Research Program through the National Research Foundation of Korea (NRF) funded by the Ministry of Education (RS-2023-00245035).

\appendix

\section{How to determine $\phi_j$ and $\tau_j$ for $j=1,2$}\label{AppA}
From eqs.~(\ref{w1w2}) and (\ref{wphitau}), we obtain
\begin{align}
\begin{split}\label{eqphitau}
e^{2\phi_1}=&\,w_1\bar{w}_1=\epsilon^2+l^2-t^2-2i\epsilon t,\\
e^{2i\tau_1}=&\,\frac{w_1}{\bar{w}_1}=\frac{-\epsilon^2+l^2-t^2-2i\epsilon l}{\epsilon^2+(l+t)^2},\\
e^{2\phi_2}=&\,w_2\bar{w}_2=\epsilon^2+l^2-t^2+2i\epsilon t,\\
e^{2i\tau_2}=&\,\frac{w_2}{\bar{w}_2}=\frac{-\epsilon^2+l^2-t^2+2i\epsilon l}{\epsilon^2+(l+t)^2}.
\end{split}
\end{align}
Note that $\phi_j $ and $\tau_j$ are real if $t=0$.
One can find solutions of the above equations with 
\begin{align}\label{condition2}
-\pi/2\le\Im [\phi_{j}]\le\pi/2,& \;\;\; \pi/2\le\Im [i\tau_{j}]\le3\pi/2,
\end{align}
which satisfy eq.~(\ref{condition}). However, these solutions may not satisfy eq.~(\ref{wphitau}) because eq.~(\ref{eqphitau}) is invariant under 
\begin{align}\label{reflection}
w_j\to-w_j, \;\;\; \bar{w}_j\to-\bar{w}_j.
\end{align}
 If they do not satisfy  eq.~(\ref{wphitau}), we can construct correct ones with eq.~(\ref{condition}) from the solutions with eq.~(\ref{condition2}) by one of the following shifts for eq.~(\ref{reflection}):
\begin{itemize}
\item If $0\le\Im [\phi_j+i\tau_j]\le\pi, -\pi\le\Im [\phi_j-i\tau_j]\le0$, 

\noindent
shift $\phi_j\to\phi_j, \tau_j\to\tau_j+\pi$.
\item If $\pi\le\Im [\phi_j+i\tau_j]\le2\pi, -2\pi\le\Im [\phi_j-i\tau_j]\le-\pi$, 

\noindent
shift $\phi_j\to\phi_j, \tau_j\to\tau_j-\pi$.
\item If $0\le\Im [\phi_j+i\tau_j]\le\pi, -2\pi\le\Im [\phi_j-i\tau_j]\le-\pi$, 

\noindent
shift $\phi_j\to\phi_j+i\pi, \tau_j\to\tau_j$.
\item If $\pi\le\Im [\phi_j+i\tau_j]\le2\pi, -\pi\le\Im [\phi_j-i\tau_j]\le0$, 

\noindent
shift $\phi_j\to\phi_j-i\pi, \tau_j\to\tau_j$.
\end{itemize}

\section{Numerical plots of eq.~(\ref{CoshPhin-CosTaun})}\label{AppB}
In this appendix, we numerically check the $n$-dependence of eq.~(\ref{CoshPhin-CosTaun}) in the late time limit to see when eq.~(\ref{poleEq}) holds. FIG.~\ref{Fig1} shows numerical plots of eq.~(\ref{CoshPhin-CosTaun}) with $t=10, l=1, \epsilon=0.1$ . From the upper figure (\ref{fig:PlotPolev1}), one can see that $\cosh (\Delta \phi/n)-\cos (\Delta \tau/n)$ is zero when $n\sim \pm1, \pm\frac{1}{2}, \pm\frac{1}{3},\cdots$, which shows that eq.~(\ref{poleEq}) holds if $\sin (\pi/n)\sim0$. The lower figure (\ref{fig:PlotPolev2}) is an enlarged plot of eq.~(\ref{CoshPhin-CosTaun}) around $n=1$. We can confirm that the value of $n$ for which $\cosh (\Delta \phi/n)-\cos (\Delta \tau/n)=0$ holds is slightly different from $n=1$. This difference is due to $O(t^{-2})$ terms in eq.~(\ref{CoshPhin-CosTaun}).

\begin{figure}[h]
  \centering
       \begin{subfigure}[b]{0.5\textwidth}
         \centering
         \includegraphics[width=\textwidth]{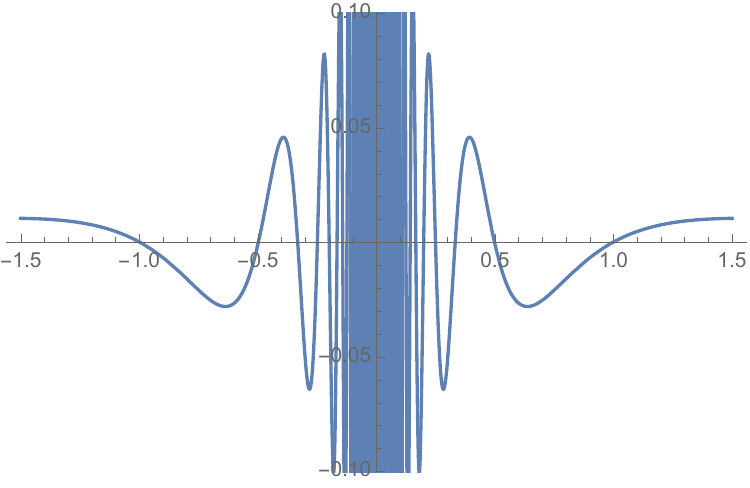}
       \put(5,60){$n$}
    \put(-170,145){$\cosh (\Delta \phi/n)-\cos (\Delta \tau/n)$}
      \caption{$n$-dependence of $\cosh (\Delta \phi/n)-\cos (\Delta \tau/n)$.}\label{fig:PlotPolev1}
        \end{subfigure}
     \begin{subfigure}[b]{0.5\textwidth}
         \centering
         \includegraphics[width=\textwidth]{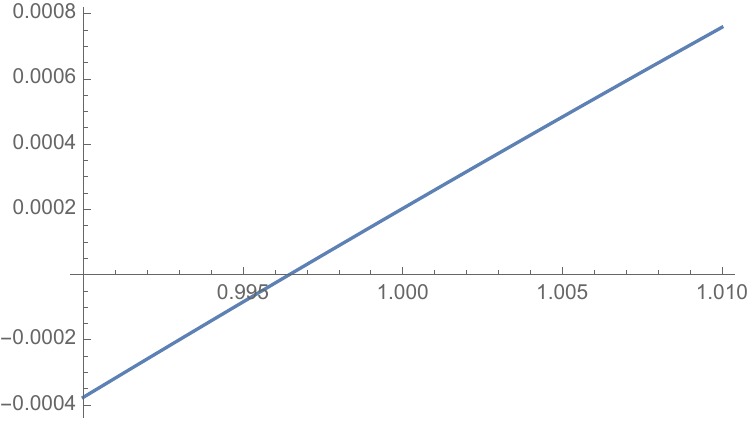}
          \put(5,33){$n$}
     \put(-260,125){$\cosh (\Delta \phi/n)-\cos (\Delta \tau/n)$}
         \caption{Enlarged plot around $n=1$.}\label{fig:PlotPolev2}
     \end{subfigure}
     \caption{Numerical plots of $\cosh (\Delta \phi/n)-\cos (\Delta \tau/n)$ in eq.~(\ref{CoshPhin-CosTaun}) with $t=10, l=1, \epsilon=0.1$ .} 
     \label{Fig1}
\end{figure}

\bibliography{Refs}
\bibliographystyle{utphys}

\end{document}